\documentclass[intlimits,twoside,a4paper]{article}

\usepackage[T2A]{fontenc} 
\usepackage[greek, ukrainian,english]{babel}
\usepackage[cp1251]{inputenc}
\usepackage{cmpj3}

\usepackage{amsmath}
\usepackage{empheq}
\usepackage{pstricks}
\usepackage{mathrsfs}
\usepackage{mathtools}
\usepackage{cancel}
\usepackage{accents}
\usepackage{dsfont}
\usepackage[normalem]{ulem}
\usepackage[caption = false]{subfig}

\usepackage{upgreek}
\def\coppa{{\fontencoding{LGR}\fontfamily{cmr}\selectfont\textqoppa}}

 \DeclareRobustCommand{\augiefamily}{%
 	\fontfamily{augie}\fontseries{m}\fontshape{n}\selectfont}
 \DeclareTextFontCommand{\textaugie}{\augiefamily}

\def\be{\begin{equation}}
	\def\ee{\end{equation}}
\def\bea{\begin{eqnarray}}
	\def\eea{\end{eqnarray}}
\def\bpar{\left(\!\!\begin{array}}
	\def\epar{\end{array}\!\!\right)}
\def\bdar{\left|\!\!\begin{array}}
	\def\edar{\end{array}\!\!\right|}
\def\barr{\begin{array}}
	\def\earr{\end{array}}
\def\btab{\begin{tabular}}
	\def\etab{\end{tabular}}

\def\vac#1{{\bf #1}}

\def\<{\langle}
\def\>{\rangle}

\def\qq{{\hbox{\foreignlanguage{greek}{\coppa}}}}
\def\qqq{{\hbox{\foreignlanguage{greek}{\footnotesize\coppa}}}}

\issue{2024}{27}{1}{13603}
\doinumber{10.5488/CMP.27.13603}

\title[Finite-size scaling in free boundary conditions above the upper critical dimension]{When correlations exceed system size: finite-size scaling in free boundary conditions above the upper critical dimension}

\author[Yu. Honchar, B. Berche, Yu. Holovatch, R. Kenna]{{Yu. Honchar}\orcid{0000-0003-2660-4593} \refaddr{inst1,inst2,inst4}
    , {B. Berche}\orcid{0000-0002-4254-807X}\refaddr{inst3,inst4}
    , {Yu. Holovatch}\orcid{0000-0002-1125-2532}\refaddr{inst1,inst2,inst4,inst5}
    , \framebox{R. Kenna}\orcid{0000-0001-9990-4277}\refaddr{inst2,inst4}%
    }
		\addresses{
		\addr{inst1}{{Institute for Condensed Matter Physics of the National Academy of Sciences of Ukraine}, 
		 79011 Lviv, Ukraine}
		\addr{inst2}{{Centre for Fluids and Complex Systems, Coventry University}, 
			Coventry CV1 5FB,
			{UK}}
		\addr{inst3}{{Laboratoire de Physique et Chimie Th\'eoriques, Universit\'e de Lorraine - CNRS, Nancy
Cedex, France}}
		\addr{inst4}{{$\mathbb{L}^4$ Collaboration and Doctoral College for the Statistical Physics of Complex Systems}, 
			{Lviv-Leipzig-Lorraine-Coventry}, 
			{Europe}}
   \addr{inst5}{{Complexity Science Hub Vienna, 1080 Vienna, Austria}}
		}

\date{Received September 15, 2023, in final form November 20, 2023}

\begin{document}

		\maketitle
		\begin{abstract}

   We progress finite-size scaling in systems with free boundary conditions above their upper critical dimension, where in the thermodynamic limit critical scaling is described by  mean-field theory.
   Recent works show that the correlation length is not bound by the system's physical size, a belief that long held sway. 
   Instead, two scaling regimes can be observed --- at the critical and pseudo-critical temperatures.
   We demonstrate that both are manifest for free boundaries. 
   We use numerical simulations of the $d=5$ Ising model to analyse the magnetization, susceptibility, {{magnetization Fourier modes and the partition function zeros.}}
   While some of the response functions hide the dual finite-size scaling, the precision enabled by the analysis of Lee--Yang zeros allows this be brought to the fore. 
   In particular, finite-size scaling of leading zeros at the pseudo-critical point confirms recent predictions coming from correlations exceeding the system size.
   This paper is dedicated to Jaroslav Ilnytskyi on the occasion of his 60th birthday.

		\end{abstract}

Over ten years ago {{our team}} published a paper on finite-size scaling (FSS) in systems in high dimensions which triggered a considerable reaction in the community \cite{pbc3}. 
FSS in its standard form was for fifty years expected to fail in high dimensions and our paper introduced a new form that claimed to rescue it.
However, subtleties exist --- one has to carefully differentiate between boundary conditions and precisely where criticality is measured.
A core tenet of our arguments was that universality, and hence the correct form of FSS resided at pseudo-criticality and not  at criticality.
{{At the time,}} FSS for free boundary conditions (FBCs) at the critical point was most challenging. 
At that point, and in the absence of a more complete theory, results were unconvincing  for they appeared neither to fit with standard FSS nor with the new regime.
One paper inspired by our work was that of Lundow and Markstr{\"{o}}m, {{whom we refer to as L\&M}} \cite{fbc1}.
The paper focuses entirely on FBCs at the critical (not pseudo-critical) point and uses impressive numerical power to argue that standard FSS holds there.
In the meantime, significant theoretical developments have occurred in the form 
of~references~\cite{Q1,percolation1,fbc3,Q,percolation3} 
and we refer the reader to reference~\cite{SciPost} for a new theory as well as an extended and detailed review.
The new developments come under the broad name of {Q theory, or simply Q,} with the new finite-size scaling prescription called QFSS.

Here we revisit FSS for FBCs at both the critical and pseudo-critical points.
We argue that what appeared in L\&M as ``standard'' FSS is in fact a form of QFSS that falls within Q theory --- it is FSS applied to the Gaussian model instead of to Landau mean-field theory.
To differentiate between standard FSS applied to Landau theory and Q theory applied to the Gaussian fixed point, we revisit the magnetization (which L\&M  also looked at) as well as Fourier modes. We also examine Lee--Yang zeros at the pseudo-critical and critical temperatures.
In terms of the observables used in the study of phase transitions and critical phenomena, zeros are unique in that they are not moments of the free energy. 
Together, these provide evidence that validates Q theory for {{FBCs}} in both regimes.

The structure of the paper is {{as follows. 
In}}  the first section we look at the background of our research, and discuss previous results on the topic; 
in section 2 we present {{the results of}} our numerical simulations and the finite-size scaling of the {{various}} observables; 
section 3 {{documents}} the search and then the scaling of the partition function zeros; 
results and conclusions are {{laid}} out in the last section. 
It is our special pleasure to dedicate this paper to Jaroslav Ilnytskyi, a colleague and a good friend of ours, whose impressive work on numerical simulations in condensed matter physics, and in particular on universality at critical points  
{{\cite{Iln,Iln2,Iln3} has inspired us for many years and continues to do so today.}}

 \section{Introduction}\label{I} 

The subject of phase transitions and critical phenomena is key to unraveling how matter undergoes transformations \cite{books0}.
A first approximation of this is mean-field theory, a subject that is now over 100 years old~\cite{Tours}.
Once a certain space dimension --- called the upper critical $d_{\rm c}$ --- is surpassed, the number of interactions is so large that the system's behavior is effectively averaged, and mean-field theory holds sway. 
While valid for infinite systems, this is often also considered to be the case (with suitable adaption) for finite systems too. 
However, it was recently exposed that this is not always the case and one has to take into account the crucial  role played by correlation length. 
Until recently this was widely believed to scale no faster than physical size if the system is finite in extent.
This is now known not to be true and  correlations can exceed the system size. 
While significant theoretical progress has been made in verifying this picture for systems with periodic boundary conditions (PBCs), 
free boundary conditions (FBCs)  have proved more difficult.  
In this paper, we explore what happens in this lesser-known territory using computer simulations. 
We consider the Ising model, which has an upper critical dimension $d_{\rm c}=4$, on a five-dimensional hypercubic lattice. 

First, we set the notation.
For second-order phase transitions in ferromagnetic systems,  the specific heat, magnetization, susceptibility, and correlation length  near the transition temperature ${{T_c}}$, scale as
$ {{c_\infty}}(t) \sim |t|^{-\alpha}$, 
$ {{m_\infty}}(t) \sim  (-t)^{\beta}$,
$ {{m_\infty}}(h) \sim  |h|^{{1}/{\delta}}$,
$ {{\chi_\infty}}(t) \sim  |t|^{-\gamma}$,
$ {{\xi_\infty}}(t)\sim |t|^{-\nu}$,
where  $t$ is a reduced temperature, 
$h$ is a reduced magnetic field 
(each is omitted from functional arguments when it vanishes), 
{{and}} the formula for ${{m_\infty}}(t)$ holds only in the broken-symmetry phase where $t<0$,
{{and the subscript denotes the (infinite) size of the system.}}
The critical diverging singularities are replaced by finite peaks when the linear extent, $L$, of the system is finite. 
These finite peaks are centered around the so-called pseudo-critical points $T_L$.
The rounding of the susceptibility peak can be measured by its width at half-height, for example, and is governed by an exponent $\rho$: 
 $\Delta T_{\rm{rounding}} \sim L^{-\rho}$.
The distance of the pseudo-critical temperature $T_L$ from the critical one ${{T_c}}$ is described by the shift exponent $\lambda$:
$ T_L - {{T_c}} \sim L^{-\lambda}$.
Usually (but not always)
\cite{JaKe02},  
the rounding and  shifting exponents coincide and are given by 
\begin{equation}
 \rho = \lambda = \frac{1}{\nu}.
\label{usual}
\end{equation}
Below the critical dimension $d_c$ and near the critical point, the infinite-volume correlation length $\xi_\infty$ can be replaced by the system size $L$ when the latter is finite. 
This is sufficient to deliver appropriate FSS there.
{{
 Because for specific heat it leads to 
 $c \sim |t|^{-\alpha} = \xi^{\alpha/\nu}$ 
 being replaced by 
 $L^{\alpha/\nu}$.
A similar applies to other observables so that
}} 
\begin{equation}
c_L \sim L^{\frac{\alpha}{\nu}},
\quad 
m_L \sim  L^{-\frac{\beta}{\nu}},
\quad 
\chi_L \sim L^{\frac{\gamma}{\nu}}.
\label{FSS}
\end{equation}
Equations~(\ref{usual}) and (\ref{FSS}) well describe FSS below the upper critical dimension. 
When the dimensionality $d$ exceeds upper criticality  $d_c$, our current picture is  as follows~\cite{SciPost}.
The self-interacting term, which is responsible for the existence of the phase transition by spontaneous symmetry breaking (e.g., the $\phi^4$ term in the field-theoretic  equivalent to the Ising-model Hamiltonian), modifies thermodynamic functions above the critical dimension in such a  way that the above FSS formulae no longer apply.
The usual expression for the hyperscaling relation, $\nu d = 2 - \alpha$, also fails.
These circumstances can be traced back to the fact that the quartic term is coupled to  a dangerous  irrelevant  variable {{\cite{Fisher-2}}}.

Following ``surprising'' analytical~\cite{Brezin} and numerical~\cite{pbc2} observations of a superlinear (wrt $L$) correlation length, it was shown in references~\cite{Q1,ourEPL,LRI,fbc3} 
that, contrary to earlier expectations, these features of the model are also affected by such dangerous irrelevant variables, just like the properties related to free energy~\cite {BNPY}. 
This leads to the introduction of a pseudo-critical exponent $\qq= \max(1,d/d_c)$ for the FSS of the correlation length, which takes the value $d/d_c$ when $d>d_c$~\cite{pbc3,Q1,ourEPL}. 
At first sight, the  exponent $\qq$ has a similar status to $\lambda$ and $\rho$, above, in that it most directly refers to finite-size systems as opposed to the infinite-volume ones which are directly described by $\alpha$, $\beta$, $\gamma$, $\delta$, and $\nu$. 
Thus, $\qq$ may be considered a pseudo-critical exponent.
The superlinear behavior of the finite-size correlation length has been verified by Monte Carlo simulations
in references~\cite{pbc3,Q1,ourEPL,LRI,ourreview,fbc3} 
and its existence modifies the finite-size scaling relations because instead of replacing $\xi_\infty$ by $L$, one replaces it by 
\begin{equation}
 \xi_L \sim L^{\qq},
\label{Q13th}
\end{equation}
 so that
\begin{equation}
c_L \sim L^{\frac{\qqq\alpha}{\nu}},
\quad 
m_L \sim  L^{-\frac{\qqq\beta}{\nu}},
\quad 
\chi_L \sim L^{\frac{\qqq\gamma}{\nu}}.
\label{QFSS}
\end{equation}
To distinguish equation~(\ref{FSS}) from  equation~(\ref{QFSS}), we referred to the latter as QFSS~\cite{pbc3}
(the ``Q'' here and throughout referring to $\qq$, a notation suggested by Fisher in reference~\cite{ourEPL}).

The form $\xi \sim L^{d/4}$ for the correlation length of the short-range Ising-model (for which $d_c=4$) in the analytical study of reference~\cite{Brezin}  and verified numerically in 
reference~\cite{pbc2} was for periodic boundary conditions (PBCs).
That it also applies to free boundary conditions ({{FBCs}}) was suggested in reference~\cite{pbc3} and  verified in references~\cite{Q1,ourEPL,LRI,ourreview,fbc3}.
The existence of $\qq$ as a new, universal pseudo-critical exponent allows one to extend the hyperscaling relation to beyond the upper critical dimension. This general form (valid in all dimensions) is 
\begin{equation}
 \frac{\nu d}{\qq} = 2-\alpha.
\label{hyperscaling}
\end{equation}
In contrast to what was suggested above where $\qq$ appeared to be a pseudo-critical exponent,  this formula links $\qq$ to critical exponents. This promotes the new exponent to the level of critical (as opposed to pseudo-critical) exponent. Whether one considers $\qq$ to be a critical or pseudo-critical exponent is perhaps a personal preference.

To understand the origins of the exponent $\qq$, the Fourier modes associated with high-dimensional systems may be partitioned into two types~\cite{fbc3}. 
Although both are governed by the Gaussian fixed point,
so-called Q-modes are susceptible to dangerous irrelevant variables while so-called G-modes are not.
Here and throughout, following the terminology used in reference~\cite{fbc3},  ``G'' refers to  the ``pure'' Gaussian sector (unmodified by dangerous irrelevant variables) to distinguish it from the Q-sector~\cite{fbc3}.
For systems with PBCs, only the zero mode is of the Q-type; all other modes are Gaussian. 
There, as is common below $d_c$, the rounding exponent $\rho$ has the same magnitude as the shifting exponent $\lambda$, 
$ \rho = \lambda = {\qq}/{\nu}$,
ensuring that the Q-scaling window, which is centred on the pseudo-critical point, extends as far as the critical one.
In other words, QFSS applies at $T_L$ and ${{T_c}}$ in the PBC case.

For systems with free boundaries, the partitioning is somewhat more subtle.
The dominance of Q-modes at the pseudo-critical point ensures that QFSS governs finite-size effects there. 
This is where universality resides above the upper critical dimension [i.e., the QFSS formulae (\ref{QFSS}) hold both for PBCs and  {{FBCs}} there].
The rounding and shifting exponents are
\begin{equation}
 \rho = \frac{\qq}{\nu}, \quad \lambda = \frac{1}{\nu},
\label{FBCshift}
\end{equation}
and, since $\rho > \lambda$, the shifting is bigger than the rounding so that the QFSS window, centered as it is around the pseudo-critical point, does not extend to the critical point.
FSS there is governed by G-type behavior~\cite{fbc3}.

In this paper, we perform numerical simulations to test the above picture for FBCs. 
Again, we expect (as  L\&M have actually found) that G-scaling for FSS results at the critical point and Q-scaling at the pseudo-critical one. 
L\&M, however, did not look at the pseudo-critical points, so we calibrate our efforts against theirs at the critical one only. 
The best we can do is $L=50$ which compares with the impressive $L=160$ of L\&M. 
We also follow L\&M when we look at FSS of the standard observables, namely magnetization and isothermal susceptibility.
Then, we go a step further by looking at Fourier modes and partition function zeros (which L\&M did not look at). 
These lie in the complex plane of parameters entering the partition function (i.e.,  external field or temperature). 
The notion was developed by  Lee and Yang ~\cite{LY-1,LY-2}, who studied the partition function as a polynomial in a parameter related to the external magnetic field. 
Following a similar idea, Fisher suggested the study of the zeros for the temperature complex plane~\cite{Fisher1965}. 
These ideas have been termed a ``fundamental theory of phase 
transitions''~\cite{fund}.
We discuss these developments in more detail in section \ref{definitionLY}, below.

The partition function on a finite lattice is given by 
\begin{equation}
  Z_L=\sum_{E,M} p(E,M) \re^{-\beta E + h M},
\label{pf}
\end{equation}
where the configurational energy $E$ is conjugate to $\beta = 1/kT$ ($k$ is the Boltzmann factor), the configurational magnetization $M$ is conjugate to the reduced external field $h=\beta H$ and $p(E, M)$ is the density of states. 
From the fundamental theorem of algebra, this can be expressed in terms of the 
set of Lee--Yang  zeros $\{z_j\}$  as
\begin{equation}
 Z_L(h)=A(z) \prod_{j} [z-z_j(L)],
\label{fund}
\end{equation}
where $z$ is a suitable function of $h$ (e.g., $z=\exp{h}$) 
and where $A$  denotes a non-vanishing smooth function of its arguments.

Well-established standard FSS, applicable below the upper critical dimension, gives for the Lee--Yang zeros
\begin{equation}
 h_j(L) \sim L^{-\frac{\Delta}{\nu}},
\label{FSSzo}
\end{equation}
in which 
\begin{equation}
 \Delta  = \beta \delta = \beta + \gamma 
\label{Delta13b}
\end{equation}
is the gap exponent. 
These equations are the counterparts of equation~(\ref{FSS}) for the foundations of the partition function zeros.
A crucial element of {{Q theory}} as developed in reference~\cite{pbc3} is the prediction  that, for $d>d_c$, Lee--Yang zeros scale in the Q-regime (i.e., at pseudo-criticality universally and at criticality in the PBC case) as
\begin{equation}
 h_j(L) \sim L^{-\frac{\qqq \Delta}{\nu}}.
\label{FSSz}
\end{equation}
The aim of this paper is to provide a numerical check of these  scaling formulae.
For convenience, we list in table~\ref{tab:fish} various FSS predictions for various theoretical possibilities --- standard FSS from Landau theory, Gaussian theory, and QFSS.
The first Lee--Yang zeros allow us to access the G-sector [equation~(\ref{FSSzo})] through  {{FBCs}} at criticality and the Q-sector [equation~(\ref{FSSz})] through   the pseudo-critical points.

     \begin{table}[]
     	\caption{FSS (Landau and G columns) and QFSS (last column) for the variables measured in this paper ($d=5$). We expect (and find) QFSS at the pesudocritical point, standard FSS applied to Gaussian modes at the critical point, and standard FSS which is of relevance for Landau values not to feature at all.}
     	\label{tab:fish}
         \centering
         \begin{tabular}{c|lll}
         \hline
             Quantity & Landau & G & Q \hbox{\vrule depth8pt height 12pt width 0pt}  \\ \hline\\
              $m$ 
                      & $L^{-{\beta}/{\nu}} = L^{-1}$ 
                        & $L^{-{\beta_G}/{\nu}} = L^{-{3}/{2}}$
                          & $L^{-{\beta_Q}/{\nu}} = L^{-{\qqq \beta}/{\nu}} = L^{-{5}/{4}}$ \\
              $\chi$ & $L^{{\gamma}/{\nu}} = L^{2}$ 
                        &  $L^{{\gamma_G}/{\nu}} =L^{2}$
                          & $L^{{\gamma_Q}/{\nu}} =L^{{\qqq \gamma}/{\nu}} = L^{{5}/{2}}$ \\
              $c$       & $L^{{\alpha}/{\nu}} = L^{0}$ 
                          &  $L^{{\alpha_G}/{\nu}} =L^{0}$
                            & $L^{{\alpha_Q}/{\nu}}=L^{{\qqq \alpha}/{\nu}} = L^{0}$ \\
              $h_1$     & $L^{{-\Delta}/{\nu}} = L^{-{3}/{2}}$ 
                          & $L^{{-\Delta_G}/{\nu}} = L^{-{3}/{2}}$
                            & $L^{{-\Delta_Q}/{\nu}} =L^{-{\qqq \Delta}/{\nu}} = L^{-{15}/{8}}$ \\
              \\ \hline
         \end{tabular}

     \end{table}

We will see that, as emphasized in reference~\cite{fbc3} for the magnetization, despite mean-field exponents being accurate in infinite volume, the extension of Landau theory to finite systems
{{does not manifest physically through standard FSS.}}
Instead, FSS is either of the G- or Q-types consistent with the theory developed in  references~\cite{pbc3,Q1,ourEPL,LRI,ourreview,fbc3}.

A significance of this is that pure G-scaling  was hitherto considered as merely a ``stepping stone'' to matching the finite size with the Landau mean-field exponents.
Here, we see that the G-sector has direct physical manifestation in finite volume.
Moreover, while FSS and hyperscaling were widely considered inapplicable above the upper critical dimension, here we again demonstrate that they hold in Q-modified form.
Thus, the main message of the paper is that universality holds in high dimensions, but at pseudo-criticality rather than at criticality;
this also provides a fundamental insight into how criticality at infinite volume should be approached from finite size.

 \section{Numerical simulations}

 {{Lundow and Markstr{\"{o}}m}} conducted large-scale simulations of the $d=5$ Ising model with  {{FBCs}}~\cite{fbc1,fbc2,boundary2}.
 The magnetization at $T_c$, computed using space averaging over all sites, was clearly supportive of  ${{m_L(T_c)}} \sim L^{-1.5}$. 
 (We note that this result was not determined through a free FSS fit to $\beta / \nu$ and instead expected values were imposed.)
 Table~\ref{tab:fish} suggests  this is indicative of Gaussian (rather than Landau)  FSS:  $L^{{{-\beta/\nu}}} =  L^{-3/2}$. 

 Simulations from reference~\cite{pbc3}, conducted by some of the current  authors, had previously yielded different results for the same quantity: $m_L({{T_c}}) \sim L^{-1.7}$.
 This result does not support any of the theoretical suggestions of table~\ref{tab:fish}.
 Reference~\cite{pbc3} also looked at the pseudo-critical point and found {{$m_L(T_L) \sim L^{-1.49}$.}}   Contrary to theory, this aligns with G-scaling. 
 At the time we suspected that these results --- not supportive of any theory --- were a product of the lattices used being not genuinely representative of five dimensions. 
 In particular, sites at the lattice boundaries do not have the same number of neighbours as sites in the interior of the lattice. 
 So, in reference~\cite{pbc3}, the authors {{took}} a step to minimize the role of lower-dimensional sub-manifolds in FBC systems. They computed the  ``core" magnetization, averaging only over the central $(L/2)^d$ sites. 
 For the critical point, this yielded results of $m_L^{\rm core}(T_c) \sim L^{-1.57}$, possibly in line with G-scaling at $d=5$. 
 For the pseudo-critical point, the result was {{$m_L^{\ \!\rm core}(T_L) \sim L^{-1.27}$}}, potentially aligned with Q-scaling. 
 
 There is a clear discrepancy between {{L\&M}} results in references~\cite{fbc1,fbc2} at $T_c$ and those of our team~\cite{pbc3} when all sites are used to perform the space average. The agreement is much better when only core sites are taken into account in the space average. This suggests that the previous results of reference~\cite{pbc3} are strongly affected by crossovers due to the boundary effects, while numerics in references~\cite{fbc1,fbc2} are more reliable because of the huge sizes reached.

{{Here,}} we  provide numerical simulations of the five-dimensional Ising model with free boundary conditions. We  analyze the scaling of magnetization, 
isothermal susceptibility, and magnetization Fourier modes, as well as use the analysis of the partition function zeros. 

  In order to investigate the critical behavior of the finite size $d=5$ Ising model,  we performed Monte Carlo simulations using the Wolff algorithm \cite{Wolff}. This is a cluster algorithm that updates the lattice by picking a random node and adding its neighbors to the cluster if they are of the same spin sign with probability $p = 1 - \re^{-\beta\Delta E}$, { where $\Delta E$ is a change of internal system energy if one spin is flipped} \cite{FK-1,FK-2,FS,FS-2}. In this way at least one node is flipped in every simulation step, and at zero temperature the whole lattice is added to the cluster.

  We found that starting from an ordered state of the lattice (where all spins are pointing in the same direction) we complete the thermalization process much faster than when starting with the disordered state of the lattice (where all spins are oriented randomly). 
  This effect is  more stark for larger lattice sizes. 
  The reason for this is clear --- for cluster algorithms, we need more adjacent spins of the same sign for the system to change quickly, and the disordered state does not have enough clusters to flip rapidly. 
  Therefore, we start from ordered states throughout our simulations.
    For each temperature we performed $2\times 10^6$ iterations, discarding the first 10\% of those as thermalization.

 We set boundary conditions to be free, meaning that the nodes on the edges of the lattice do not have a neighbor in one of the $d$ directions. Instead, they form a $(d-1)$-dimensional manifold.

  \subsection{Determining pseudo-criticality and its finite-size scaling}

 Besides at the critical temperature  itself, we are interested in FSS at the pseudo-critical temperature. 

 This is shifted away from the actual critical point of the infinite system and the smaller the system size the more pronounced this effect is. 
 Actually, for small lattices the system at the critical temperature is effectively disordered. 
 
\begin{figure}[hb!]
    \centering
		\subfloat[]{\includegraphics[width=0.55\linewidth]{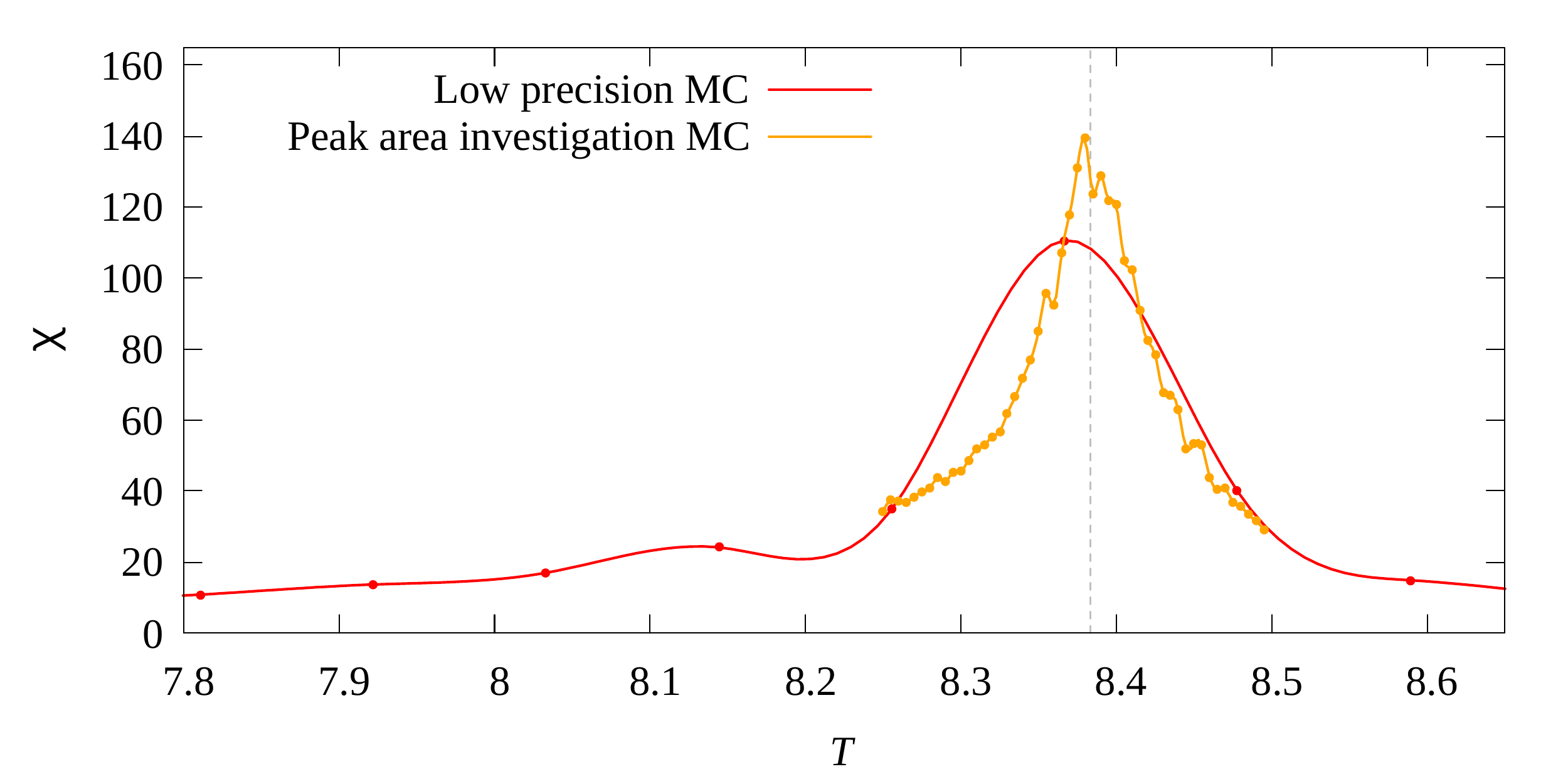} \quad}\\
		\subfloat[]{\includegraphics[width=0.55\linewidth]{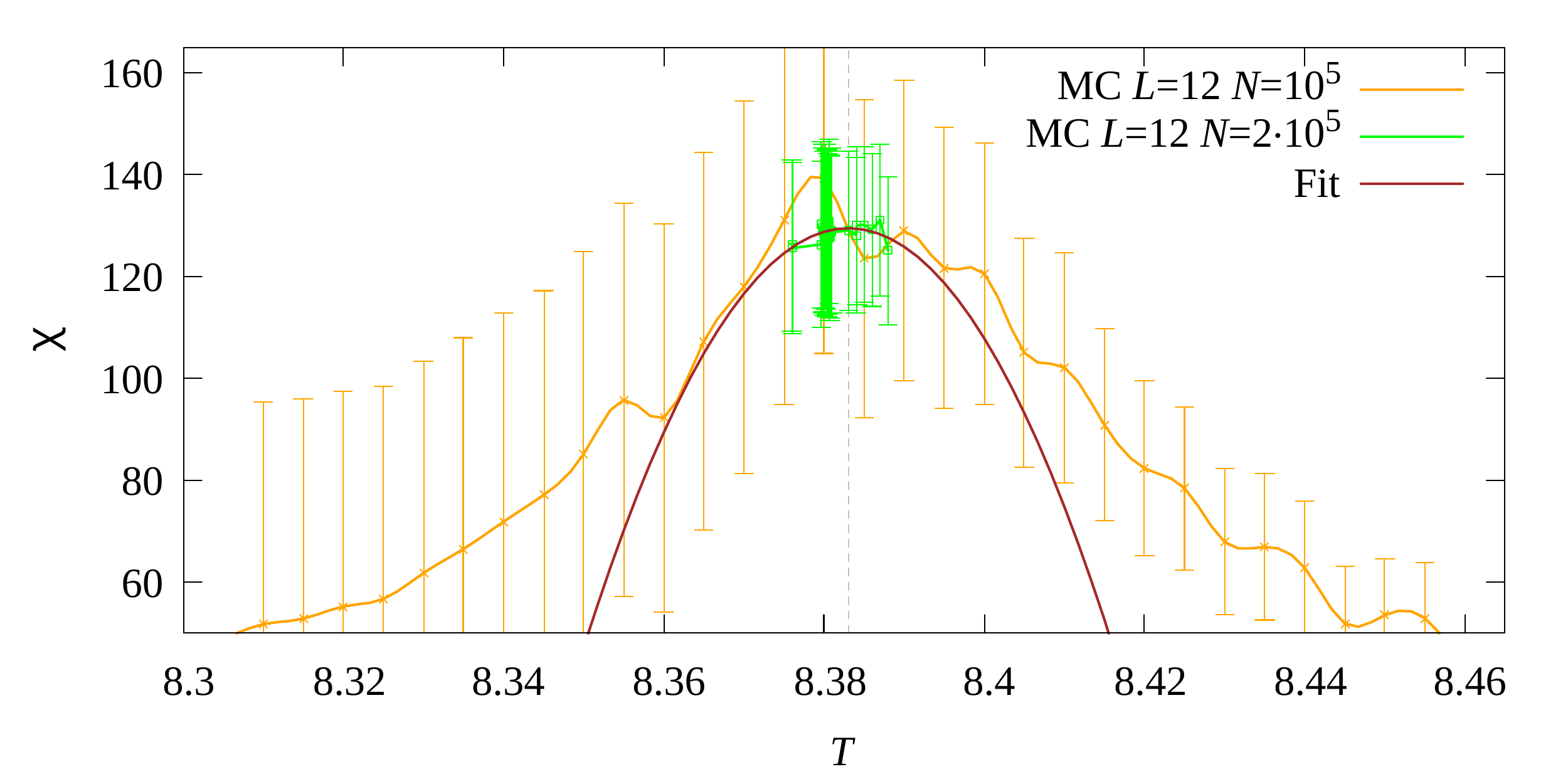} \quad}
		\caption{(Colour online) The process of estimating the pseudo-critical temperature from the susceptibility for the $d=5$ Ising model on a lattice of size $L=12$. First (a) we scan greater intervals of temperatures with relatively low precision. 
        Then, we perform this same procedure in more detail for an interval where the susceptibility is visibly higher. Finally (b) we simulate an interval of temperatures, where it seems that the peak should be located, with higher precision. We fit a polynomial through the highest points. This allows us to select an estimate for the pseudo-critical point $T_L$ (dashed vertical line). We perform final simulations at the resulting peak.
  }\label{PeakFind}
	\end{figure}

Our favored determinant of the pseudo-critical point is susceptibility. 
The algorithm to locate the pseudo-critical temperature $T_L$ is illustrated in figure~\ref{PeakFind}. 
The top part of the figure illustrates the rough values of the susceptibility for a wide range of temperature values. A peak is clearly visible and our task is to identify its location $T_L$ with some precision.   
To identify this pseudo-critical peak, then, we first scan a broad range of temperatures with low-precision simulations to find an approximate region that contains it for each lattice size. 
After that, one runs a set of high-precision simulations narrowed around this temperature.
The outcome is illustrated in the lower figure. 
Because of the fluctuations, the peak is difficult to identify manually. Thus, we use a quadratic fit in an attempt to smoothen the curve so that a local maximum can be more readily identified.
At this point, which we identify as $T_L$, we perform another high-precision simulation and use the data from it for further analysis. 

 To investigate FSS of the pseudo-critical point, we first require an estimate of the critical point itself.
 We use that ($T_c = 8.77847{{52}}$) coming from L\&M (2014) \cite{fbc1}. 
 This is the estimate of the critical point that we use throughout the paper as we also track  the behavior of various observables  at the infinite system critical point $T_c$.

\begin{figure}[ht!]
		\centering
		\includegraphics[width = 0.55 \linewidth]{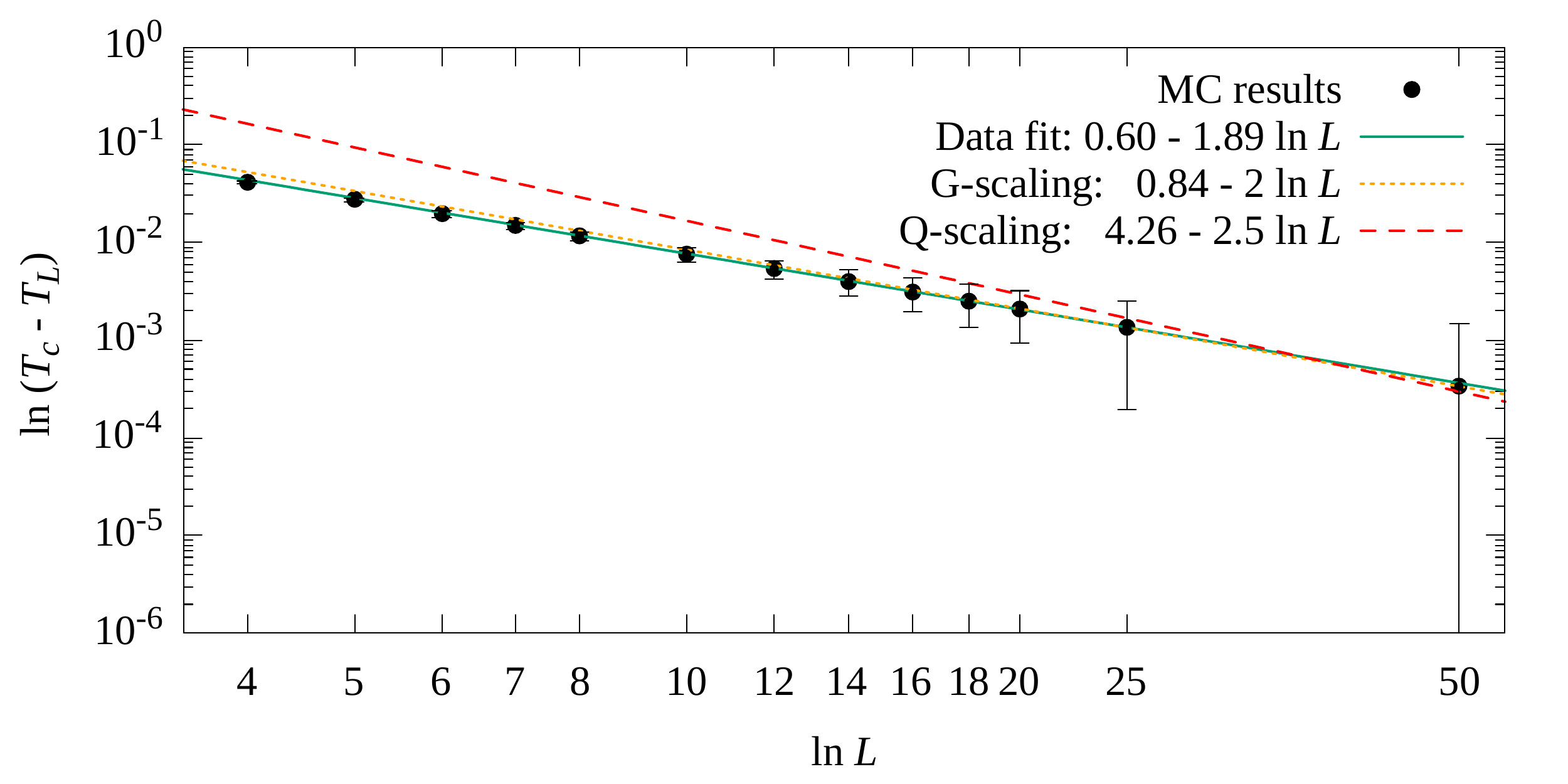}
		\caption{(Colour online) FSS data of the pseudo-critical temperature (blue line).
        For the G-scaling predicted by theory for  {{FBCs}}, $t_L$ scales as $L^{-2}$ (orange dotted line). 
        This is in marked contrast to $L^{-2.5}$ which comes from Q-scaling (red dashed line) and is relevant for PBCs. Our fit gives $L^{-1.90 \pm 0.03}$, which is G-scaling within the error bars.}
   \label{tL_scaling}
	\end{figure}   
 
  \begin{table}[b!]
  			\caption{FSS behavior for the pseudo-critical point, mean magnetization $m$, isothermal susceptibility $\chi$, magnetization Fourier modes $m(k)$, and the first Lee--Yang zero $h_1$. 
  		The first three rows record the theoretical predictions for mean-field theory (MF), Gaussian theory (G) and QFSS (Q). 
  		In  the next three rows, $T_c$ and $T_L$ refer to the results of numerical simulations at corresponding temperatures. The last lines record the results presented in this paper and exponents written in bold font are those for which there is a reasonable agreement with the theoretical predictions.}
  	\label{tab:my_label}
		\centering
		\begin{tabular}{|c|c c c c c |}
			\hline
			& $t_L$ & $m$   & $\chi$& $m(k)$ & $h_1$\\ \hline
			&&&&&\\
			MF  & $L^{-2}$  & $L^{-1}$ & $L^{2}$ &$L^{-1}$ & $L^{-3}$\\
			G-scaling  & $L^{-2}$ & $L^{-1.5}$ & $L^{2}$ & $L^{-1.5}$& $L^{-3}$ \\
			Q-scaling  & {$L^{\mathbf{-2.5}}$} & $L^{\mathbf{-1.25}}$ & $L^{\mathbf{2.5}}$ &$L^{\mathbf{-1.25}}$ & $L^{\mathbf{-3.75}}$\\ 
			at $T_c$, \cite{fbc1}  &    & $L^{-1.5}$ & $L^{2}$ &  &\\ 
            at $T_L$, \cite{pbc3} &   &$L^{-1.39}$&$L^{2}$&&$L^{-3.75}$ \\
             at $T_c$, \cite{pbc3}  &   &$L^{-1.70}$&$L^{1.71}$&&\\
            \hline
			&&&&&\\
              This paper  & $L^{1.90 \pm 0.03}$ &  &  & &  \\
			 at $T_L$, this paper  &  &  $L^{ -{\bf 1.25} \pm 0.09}$ & $L^{ 1.95 \pm 0.02}$ & $L^{-{\bf 1.26} \pm 0.05}$ & $L^{-{\bf 3.81} \pm 0.08}$  \\ 
			 at $T_c$, this paper  &   & $L^{-\bf 1.5}$ & $L^{\bf 2}$ & & $L^{-{\bf 3.16} \pm 0.48}$  \\ \hline
		\end{tabular}

	\end{table}
 
 In figure~\ref{tL_scaling} we present the FSS of the pseudo-critical temperatures $t_L=T_c-T_L$. Since this and all other
 numerical results of this paper are obtained for $d=5$, we do not mention this fact explicitly in the forthcoming 
 figure and table captions.
 Mean-field or Gaussian FSS would predict $t_L \sim L^{-1/\nu} = L^{-2}$, where $\nu$ is the correlation-lenght critical exponent. By contrast, Q-scaling predicts $t_L \sim L^{-\qqq / \nu} = L^{-2\qqq} = L^{-5/2}$. The latter was proven for the PBC case, while the former was shown for  {{FBCs}} in \cite{Rudnick}. 
 We observe $t_L\sim L^{-1.90 \pm 0.03}$ when we fit over the full range of data from $L=4 $ to $L=50$. This value is much closer to the theoretically expected G-scaling, $t_L\sim L^{-2}$, and further away from the Q-scaling prediction $t_L\sim L^{-2.5}$. 
 This is consistent with observations in the literature whereby shifting is bigger for FBCs than it is for PBCs.
 This result is recorded in table \ref{tab:my_label} as are the FSS results for other quantities. As in the case of pseudo-critical points, the table invites comparisons  with available theoretical predictions.

  \subsection[FSS of the mean magnetisation]{Finite-size scaling of the mean magnetization $m$}

	\begin{figure}[ht]
		\centering
		\includegraphics[width = 0.6\linewidth]{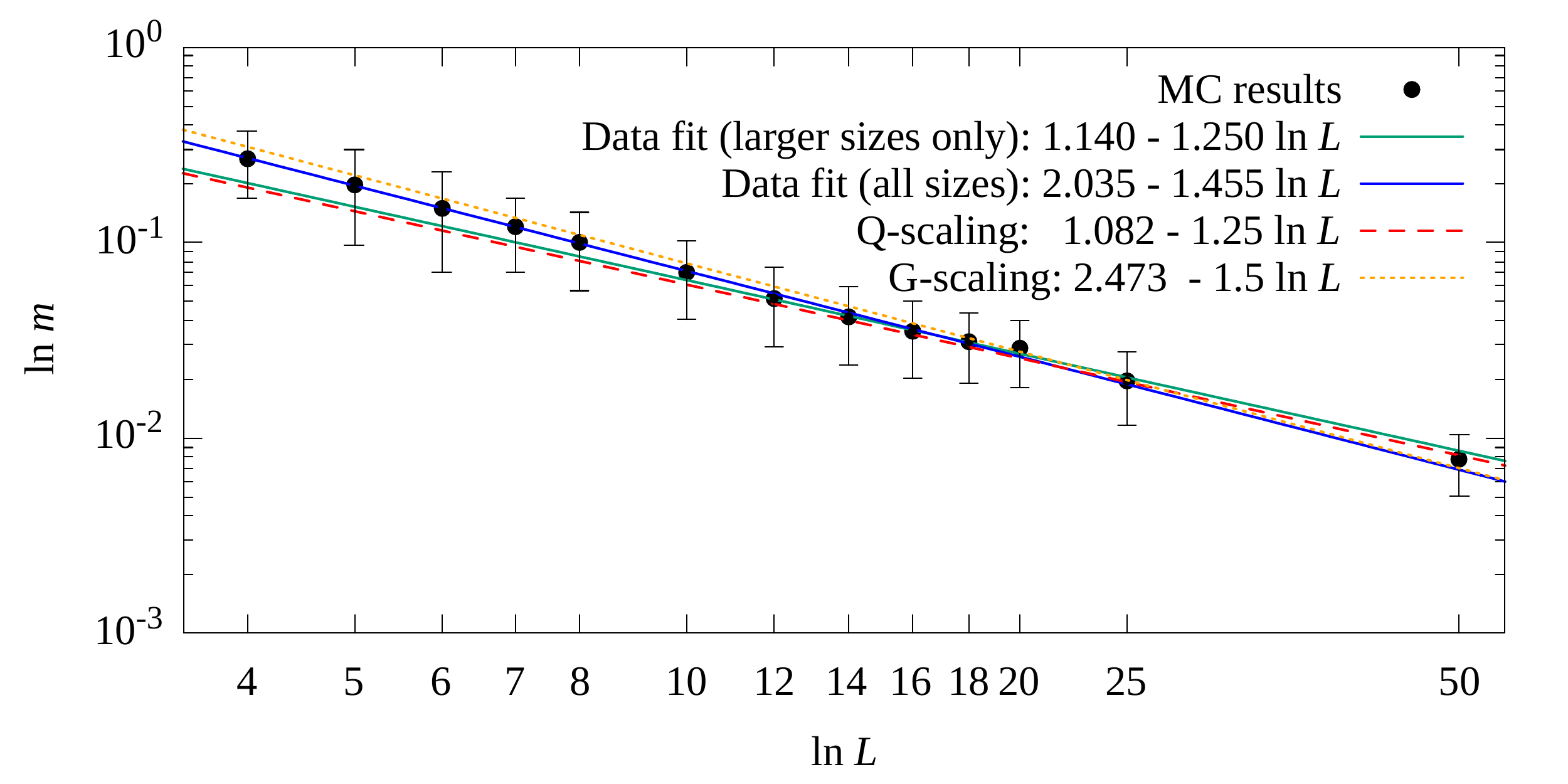}
		\caption{(Colour online) Leading behavior and first corrections to the FSS for the magnetization  at the critical temperature $T_c$.
  We plot $m\cdot L^{1.5}$ against $1/L$.}
		\label{fig:magtc}
	\end{figure}
 
	\begin{figure}[ht]
		\centering
		\includegraphics[width = 0.6\linewidth]{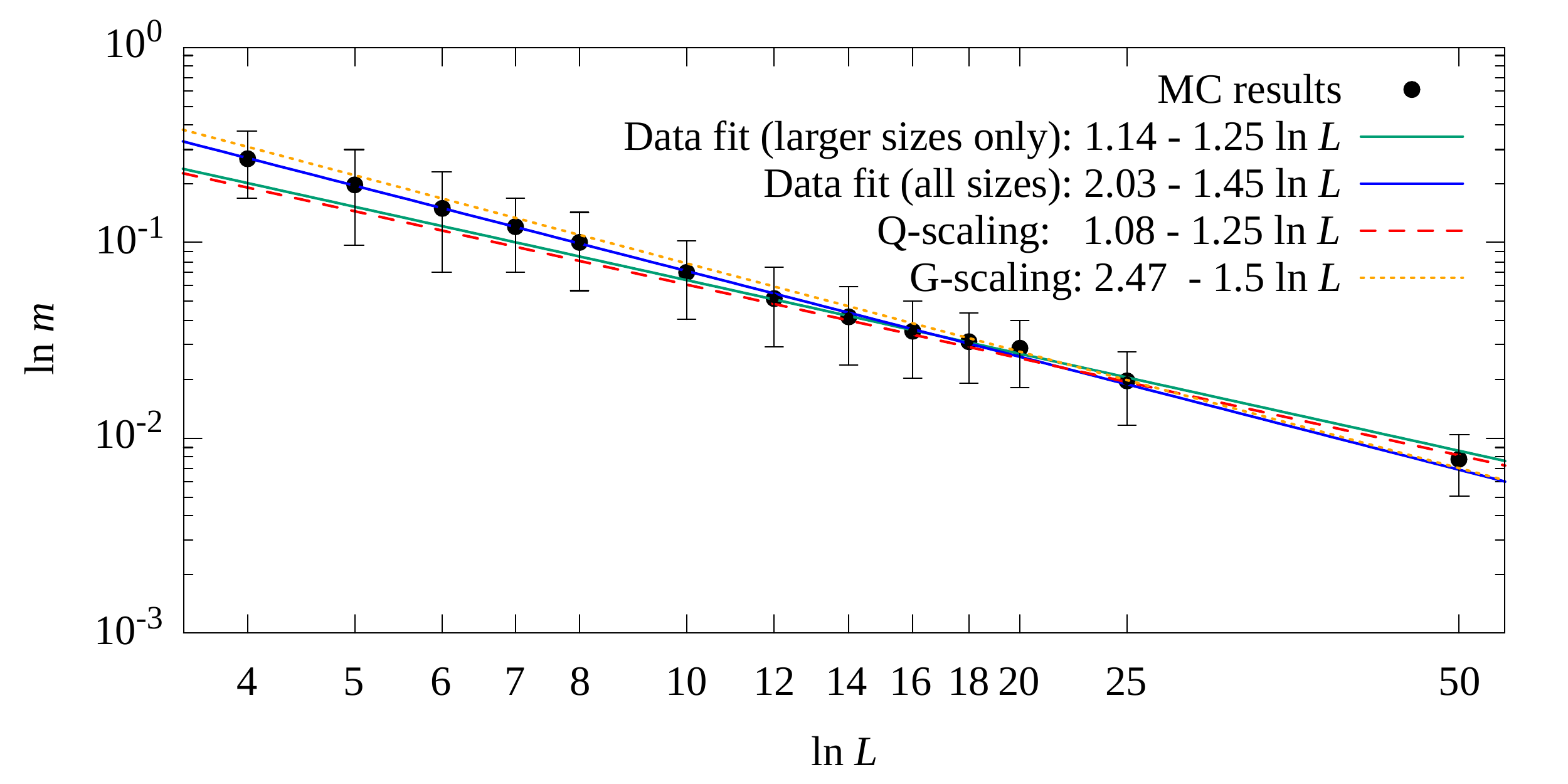}
		\caption{(Colour online) Scaling of magnetization  at the pseudo-critical temperature $T_L$. 
  The theoretical values are presented as orange dotted (G-scaling), and  red dashed (Q-scaling) lines. We see that we can discard a few smallest lattice sizes to reach the expected scaling behavior.}
		\label{fig:magtL11}
	\end{figure}

\subparagraph*{Criticality:} 

 We next analyze the FSS of mean magnetization $m$.
 Very detailed estimates for the FSS of the mean magnetization at $T_c$ were obtained by L\&M in reference~\cite{fbc1} with leading behaviors and corrections to scaling  given by:
\be
	m_L (T_c ) = 0.230 L^{-3/2} + 1.101 L^{- 5/2} - 1.63 L^{- 7/2}.
\ee
 We performed a similar analysis to compare our numerics with those of L\&M. Although extracted using far smaller lattice sizes, our numerical results deliver similar outcomes:
\be
  m_L (T_c) = 0.215 L^{-3/2} + 1.514 L^{- 5/2} -4.280 L^{- 7/2}.
\ee
 In figure~\ref{fig:magtc} we plot corrections to scaling to illustrate the  convincing nature of this fit. 
 This suggests that our results are similar to those  of L\&M and $m \sim L^{-3/2}$ as the Gaussian theory predicts.

\subparagraph*{Pseudo-criticality:} 
 
 In  figure~\ref{fig:magtL11} we show the results of the simulations for the mean magnetization $m$ at the pseudo-critical temperatures $T_L$ for different values of $L$ ranging from $L=4$ to 50.
 As was previously suggested, we expect  Q-scaling at $T_L$. At  $T_L$, we obtained $m\sim L^{-1.46 \pm 0.02}$ when the power law fit is performed over all lattice sizes. This is much closer to L\&M's G-scaling $m\sim L^{-1.5}$ than to the theoretically expected Q-scaling $m\sim L^{-1.25}$. 
 As explained in reference~\cite{pbc3}, our contention is that  the boundary lattice sites in  {{FBCs}} may be responsible for this discrepancy between the expectation and observed FSS; because the surface effects are too strong to show a properly five-dimensional criticality. 
 To illustrate this we discard the lowest sizes ($L=4..12$) from the fit, and indeed the FSS reaches the scaling $m\sim L^{-1.25 \pm 0.09}$ (green line in figure~\ref{fig:magtL}, top). 
 This suggests that for large enough lattice sizes (and discarding small lattices) the magnetization for FBCs at pseudo-criticality indeed falls within the Q-scaling regime.

  \subsection[FSS of isothermal susceptibility]{Finite-size scaling of isothermal susceptibility $\chi$}

	\begin{figure}[h!]
		\centering
		\includegraphics[width = 0.6\linewidth]{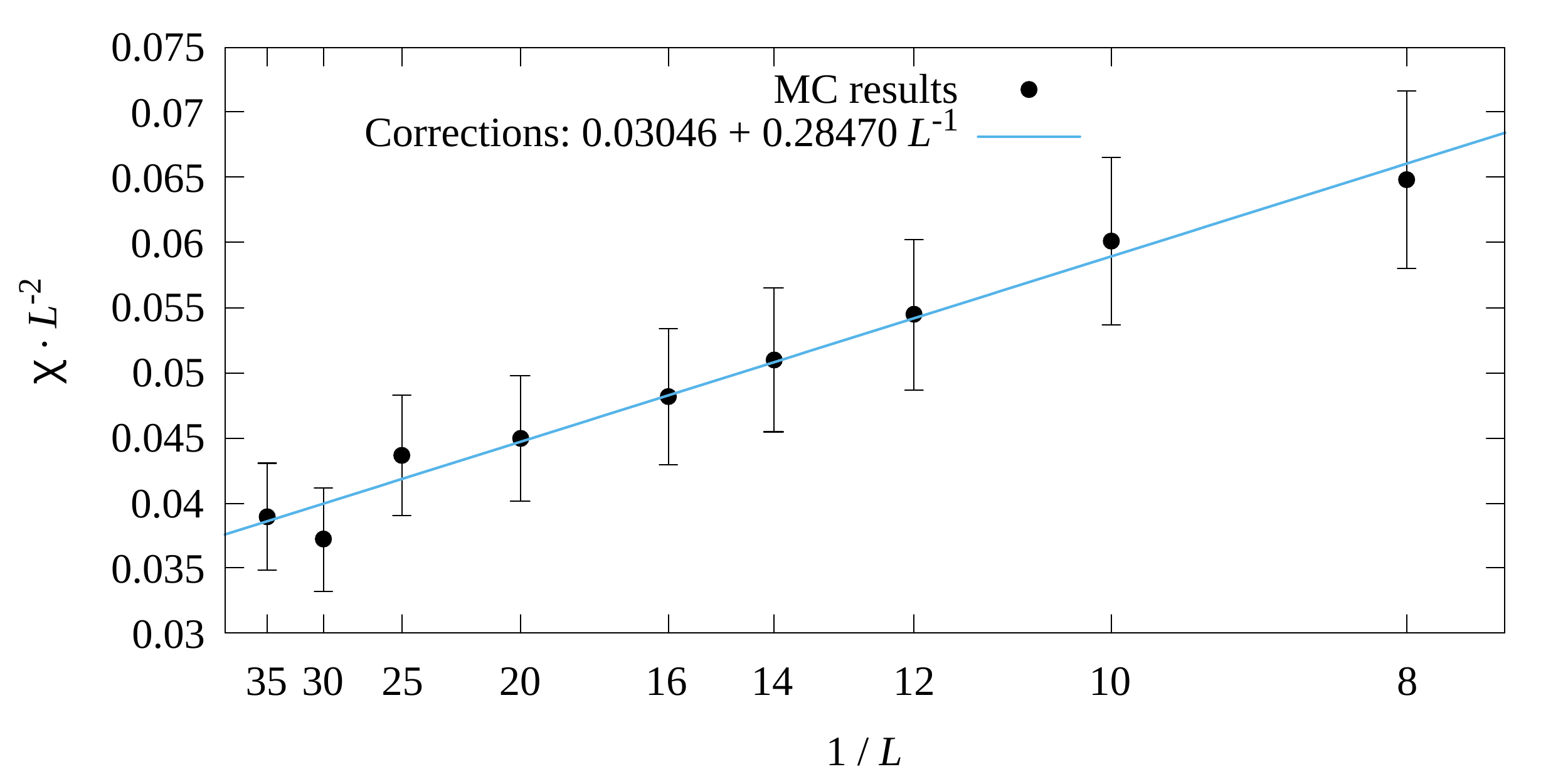}
		\caption{(Colour online) Leading behavior and first corrections to the FSS for the susceptibility at the critical temperature $T_c$. 
  For the susceptibility we plot $\chi \cdot L^{-2}$ against $1/L$.
       }
		\label{fig:magtcb}
	\end{figure}

\subparagraph*{Criticality:} 

 Again, we first compare the results of our simulations at criticality with those coming from L\&M in  reference~\cite{fbc1}.
 They assume a quadratic dependency of susceptibility on $L$ and a fit delivers
	\be
		\chi_L (T_c) = 0.817 L^2 + 0.083 L. \label{eq:chi_tc_1}
	\ee
	Our equivalent (for far smaller lattices) is
	\be 
		\chi_L (T_c) = 0.285 L^2 + 0.030 L. \label{eq:chi_tc_2}
	\ee
  The conclusion of reference~\cite{fbc1} was in favor of standard FSS scaling for the isothermal susceptibility at the critical temperature $\chi_L(T_c)\sim L^{2}$, and we can see that, at first sight, this would seem to be the case. 

 However, as table~\ref{tab:my_label} illustrates, this could be standard MF scaling or G-scaling. Our contention is that it is G and falls within the theory of Q.

\subparagraph*{Pseudo-criticality:}

\begin{figure}[ht]
		\centering
		\includegraphics[width = 0.6\linewidth]{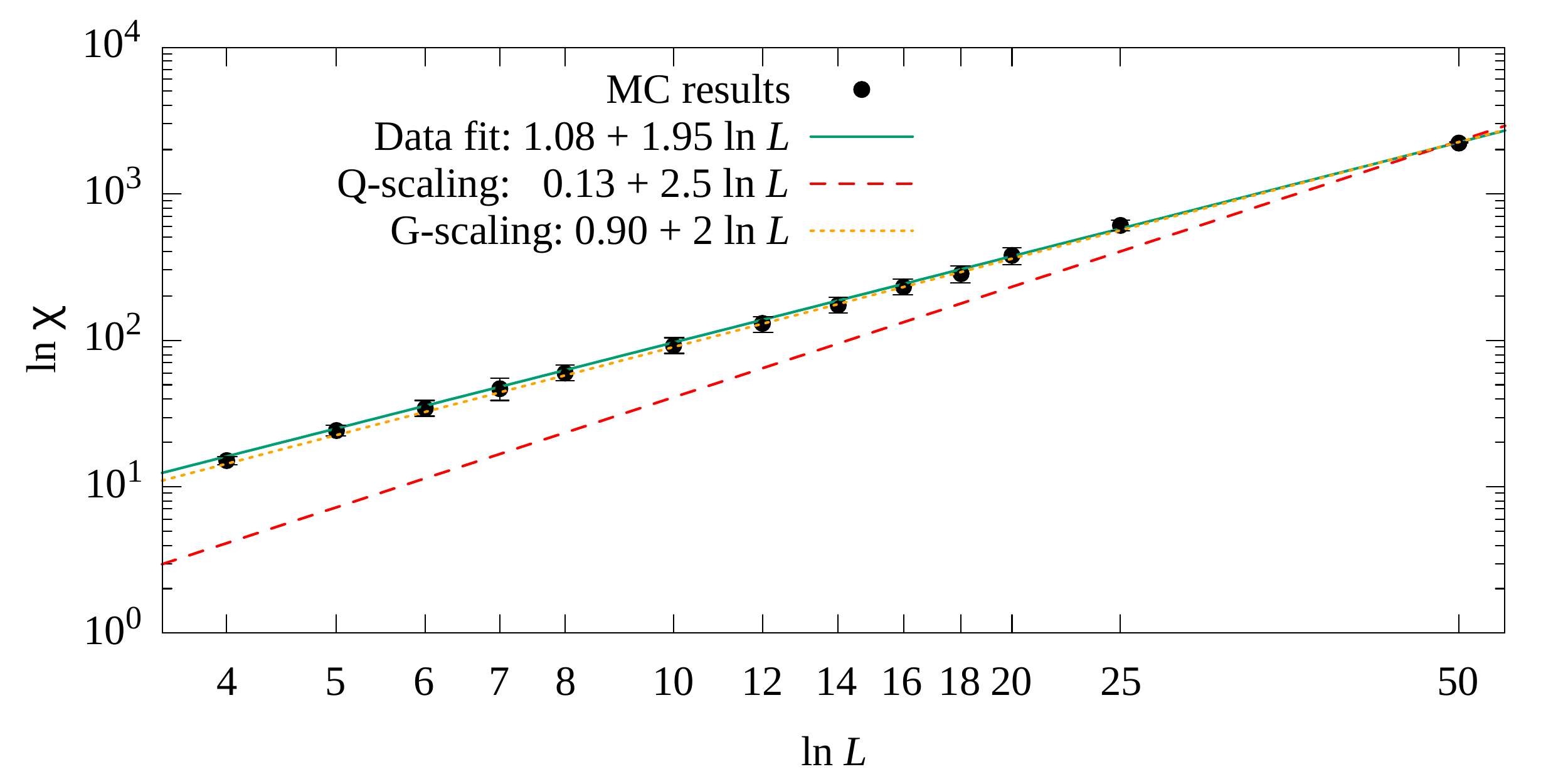}
		\caption{(Colour online) Scaling of the isothermal susceptibility at the pseudo-critical temperature $T_L$.
  The theoretical values are presented as an orange dotted line (G-scaling), and a red dashed line (Q-scaling). While  for the magnetization we were able to discard a few smallest lattice sizes to reach the expected scaling behavior, for susceptibility this is not the case and the lattice sizes are still too small to show Q-scaling.}
		\label{fig:magtL}
	\end{figure}

	Results for the FSS of isothermal susceptibility at the pseudo-critical point are shown in figure~\ref{fig:magtL}.
    At $T_L$ we obtained $\chi\sim L^{1.95 \pm 0.02}$  using all simulated lattices, in apparent agreement with G-scaling ($\chi\sim L^2$ as opposed to $\chi\sim L^{2.5}$ for Q-scaling). 
    In our fit, data for $L=4$ weigh as much as $L=50$, say. But clearly, $L=4$ is not genuinely representative of a five-dimensional system. The fact that these data lie on such a line calls into suspicion $L=50$ data as well. 
    Thus, we expect that the lattice sizes that we reach in our simulations are possibly too small to converge to Q-scaling for susceptibility. Removing a few  smaller sizes does not do the trick as it did for the magnetization. 

  \subsection[Scaling of the magetization Fourier modes]{Scaling of the magnetization Fourier modes $m(k)$}

 Since all observables are expected to be governed at $T_L$ by  Q-scaling, not G-scaling, we try other ways to  detect FSS at the pseudo-critical point for  {{FBCs}}. 
 One way to disentangle the boundary conditions  is through the Fourier transformation of the magnetization.	

 As suggested by Flores-Sola et al. in reference~\cite{fbc3} (2016), we can check the influence that the boundary conditions have on the magnetization Fourier modes and their FSS, since only some modes actually contribute to scaling for {{FBCs}}. 
 In a system with  {{FBCs,}} the Fourier modes are sine waves:
	\be
	\psi_{\vac k}=\sqrt{2/L}\prod_{\alpha=1}^d\sin k_\alpha x_\alpha,
	\ee
	with wave vectors
	$k_\alpha=n_\alpha \piup/(L+1)$, $n_\alpha=1,2,\dots,L$. Local quantities, like Ising spin $s_{\vac x_i}$, now read as
	\be
	s_{\vac x_i}=\sum_{\vac k}\tilde s_{\vac k}\psi_{\vac k}.\label{11}
	\ee
	However, in order to break the symmetry of the Ising model, we should consider the profile $\mu (\vac x)$ of the spin $s_{\vac x}$ instead:
	\be \mu (\vac x) =  \left|\left<S(\vac x) \text{sign}(M)\right>_T\right|. \ee
	Here, $\left<\dots\right>_T$ denotes temperature average. Now, Fourier transform and its inverse counterpart are:
	\bea
	\mu (\vac x)	&=& \frac{1}{(L+1)^d} \sum_{k}  \! \mu (\vac k) \ \prod_{\alpha=1}^d \! \sin ( k_\alpha\cdot {{x}_{\alpha}}) , \\
	\mu (\vac k) &=& \sum_{x=1}^N  \! \mu (\vac x) \ \prod_{\alpha=1}^d \! \sin ( k_\alpha\cdot {{x}_{\alpha}}),
	\eea
	$\vac k = \piup / (L+1) \vac{K}_L$,
	where ${x}_\alpha \in[1,L]$ represents the Cartesian coordinates of a node, $\vac{K}_L$ consists of the values in the range $(0,.., L-1)^d$.
 
	 \begin{figure}[h!]
    \centering
    \includegraphics[width=0.9\linewidth]{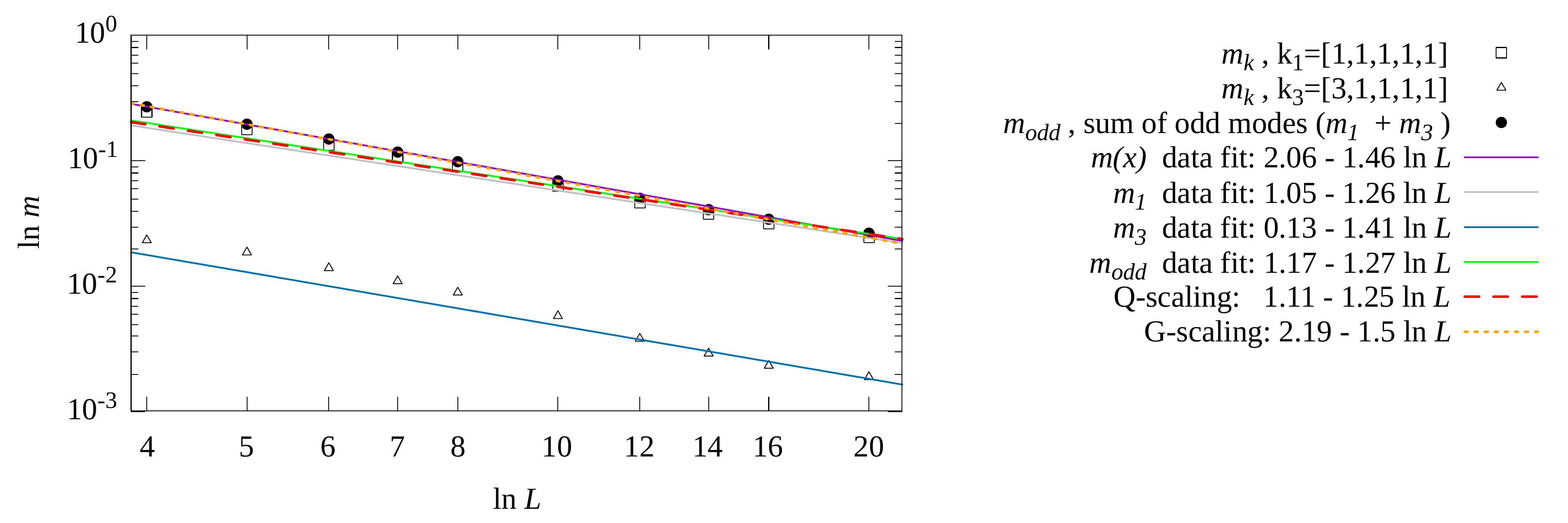}
    \caption{(Colour online) Fourier modes of the magnetization FSS in the vicinity of pseudo-critical point~$T_L$.
    As expected, even $k$ modes vanish, while odd modes provide accurate scaling.}
    \label{fig:fourier}
\end{figure}

	The mean magnetization is	
	\bea
	m &=& \frac1{L^d}\sum_x \mu(\vac x)
	= \frac1{L^d} \frac1{(L+1)^d} \sum_x\sum_k \mu(\vac k) \prod_\alpha^d \! \sin( k_\alpha\cdot {{x}_{\alpha}}) \nonumber \\
	&=& \frac1{L^d} \frac1{(L+1)^d} \sum_k\left[\sum_x \mu (\vac x) \prod_\alpha^d \! \sin( k_\alpha\cdot {{x}_{\alpha}}) \right]\cdot W_{\vac k},
	\eea
	where $W_{\vac k} = \sum_{\vac x} \prod_\alpha^d \! \sin( k_\alpha\cdot {{x}_{\alpha}})$. For even modes $\vac k$ (i.e., modes for which at least one of the $K_L$'s is even), the weights $W_{\vac k}$ are zero. These are called G-modes. Odd modes (i.e., modes for which all the $K_L$'s are odd) are called Q-modes and should contribute to the scaling. The results provided in figure~\ref{fig:fourier} and in table~\ref{tab:my_label} show that for our range of lattice sizes it is still not enough for all sites of the system to represent a five-dimensional lattice. We get $m(k=1)\sim L^{-1.26 \pm 0.05}$, which hints towards the Q-scaling, but for smaller lattice sizes again we see strong surface effects that shift the scaling towards G-scaling.  Without a definite answer, we turn to another technique that is often used in the theory of criticality: partition function zeros.

    \section{Scaling of the partition function zeros} \label{definitionLY}
 
    To summarise so far, although our results are consistent with those of L\&M at $T_c$, we have had, at best, mixed results at $T_L $, which makes it hard to disambiguate between Q suggested by 
    renormalization group theory and the G-scaling preferred by numerics.
        Therefore, we seek another way to detect whether we are in a regime with sufficiently high lattice sizes or not and we turn to partition function zeros.
    In our experience, these deliver more {{reliable}} results than those coming from thermodynamic observables.
    The latter are moments of the partition function or free energy. As such, they are sensitive to amplitudes which themselves can be variable. 
    For example, if the amplitude of correction terms for susceptibility is sufficiently large at $T_L$, the mentioned correction terms can dominate and mask the leading FSS behavior coming from power laws. In such circumstances, it might take extraordinarily large lattices for leading FSS behavior to take effect. If this happens for FBCs and not for PBCs,  it would lead to the behavior we observed. 
    Partition function zeros, on the other hand, are not subject to such amplitudes.

    For any finite system, the partition function can be written as a polynomial in terms of fugacities, and
 the partition function zeros are thus zeros in the complex plane of the appropriate variables. To analyze the phase transitions via the behavior of the zeros when approaching the critical point was first suggested by Tsung-Dao Lee and Chen-Ning Yang for complex fields \cite{LY-1,LY-2} and later was generalized for complex temperatures by Michael Fisher \cite{Fisher1965, fund}. In what follows below, we study the properties of the Lee--Yang zeros. However, for most of the models, the zeros and their critical properties can be only estimated by an extensive use of computer simulations \cite{Bena2005}.
 Analytically, solving the partition function for zeros in the complex plane is required. In particular, as the system size increases, Lee--Yang zeros exhibit a scaling behavior, accumulating near critical points in the thermodynamic limit. This scaling pattern allows for the extraction of critical exponents and the classification of phase transitions into universality classes. Understanding the scaling properties of Lee--Yang zeros provides valuable insights into critical behavior and phase transitions, enabling the characterization of macroscopic properties and underlying physical phenomena.

      \begin{figure}
    	\centering
    	\centering
     \captionsetup[subfigure]{oneside,margin={1cm,0cm}}
    	\subfloat[]{\includegraphics[width=0.5\linewidth]{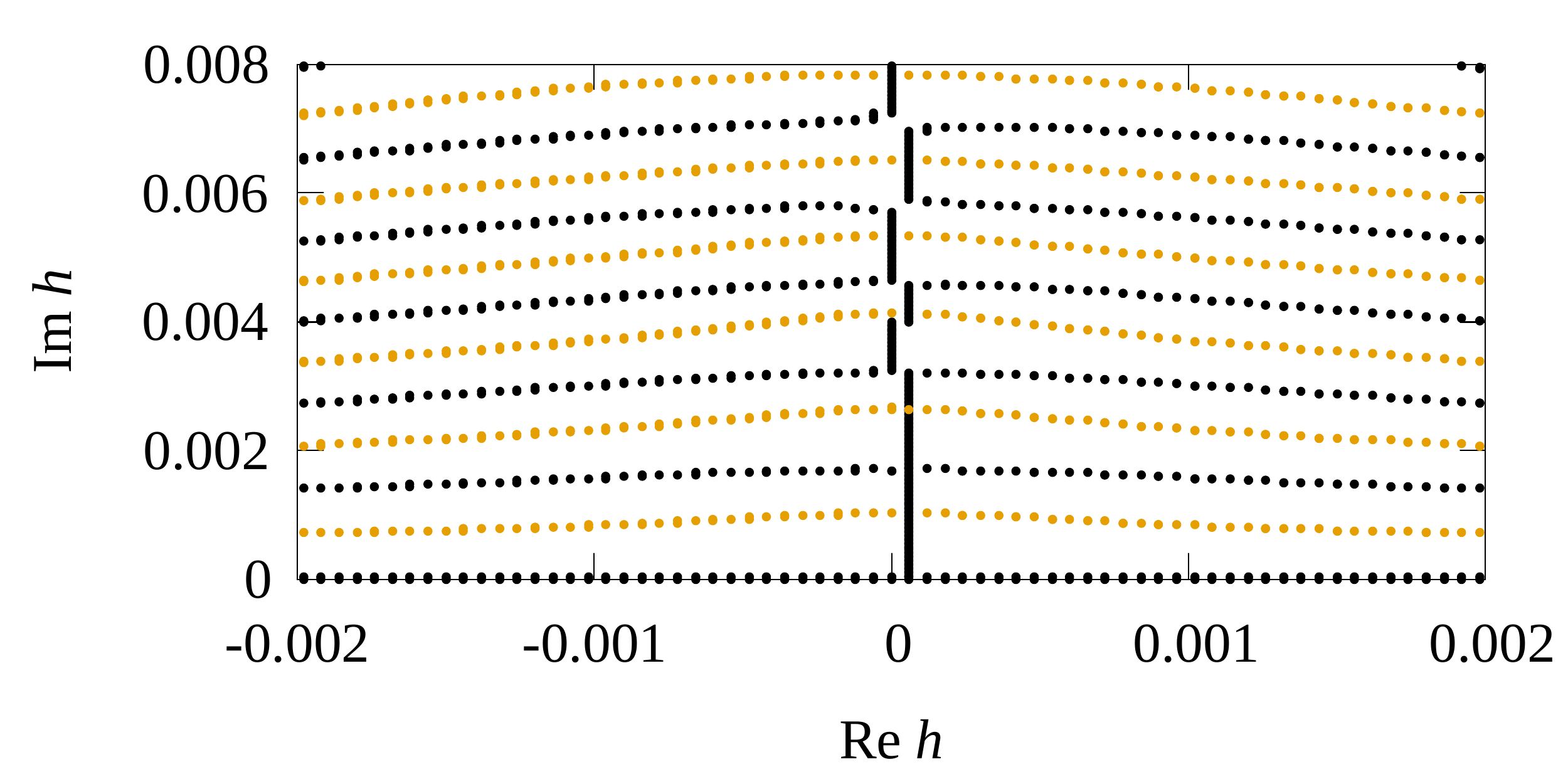}}
    	\subfloat[]{\includegraphics[width=0.5\linewidth]{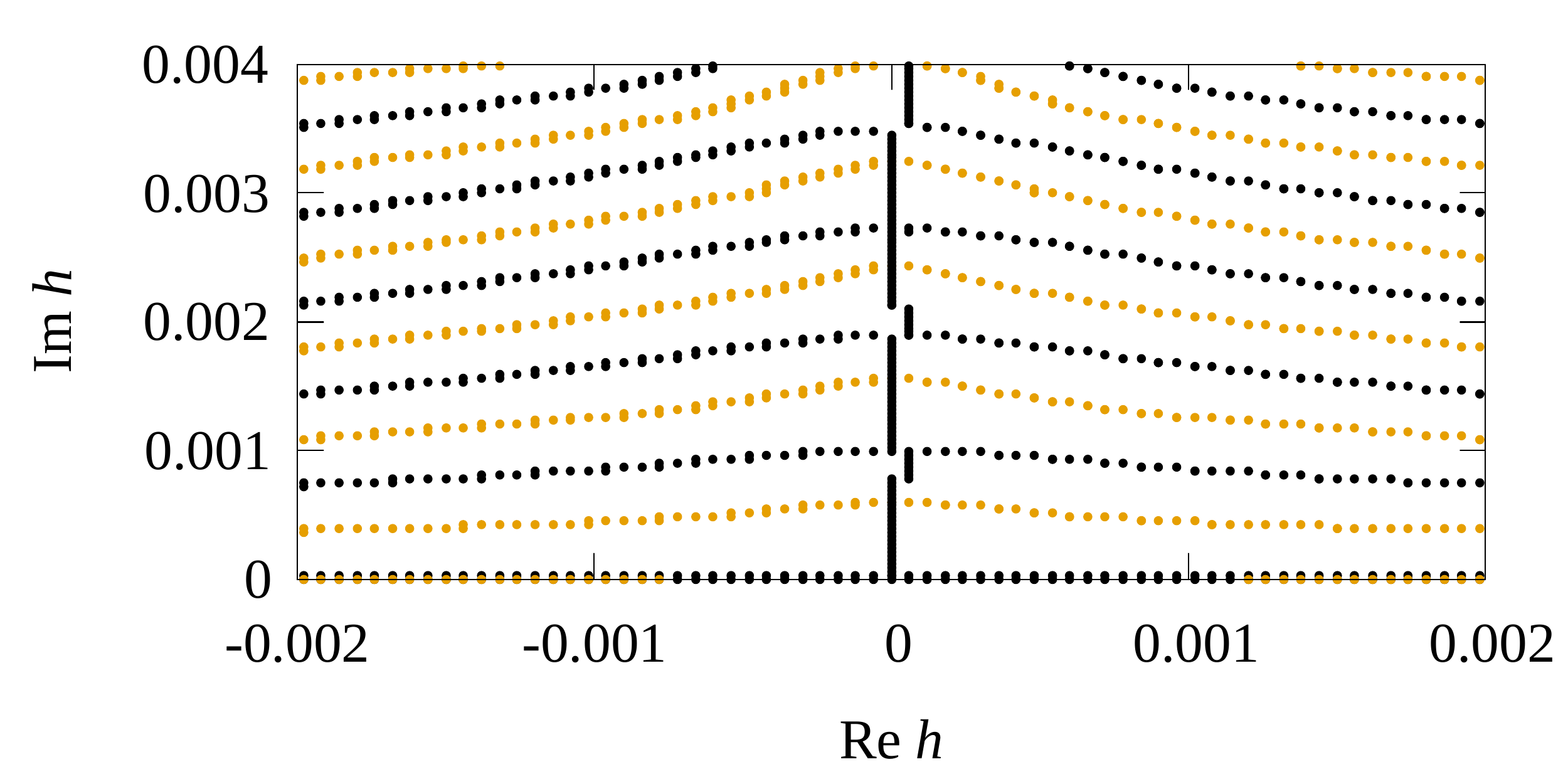}} \\
    	\subfloat[]{\includegraphics[width=0.5\linewidth]{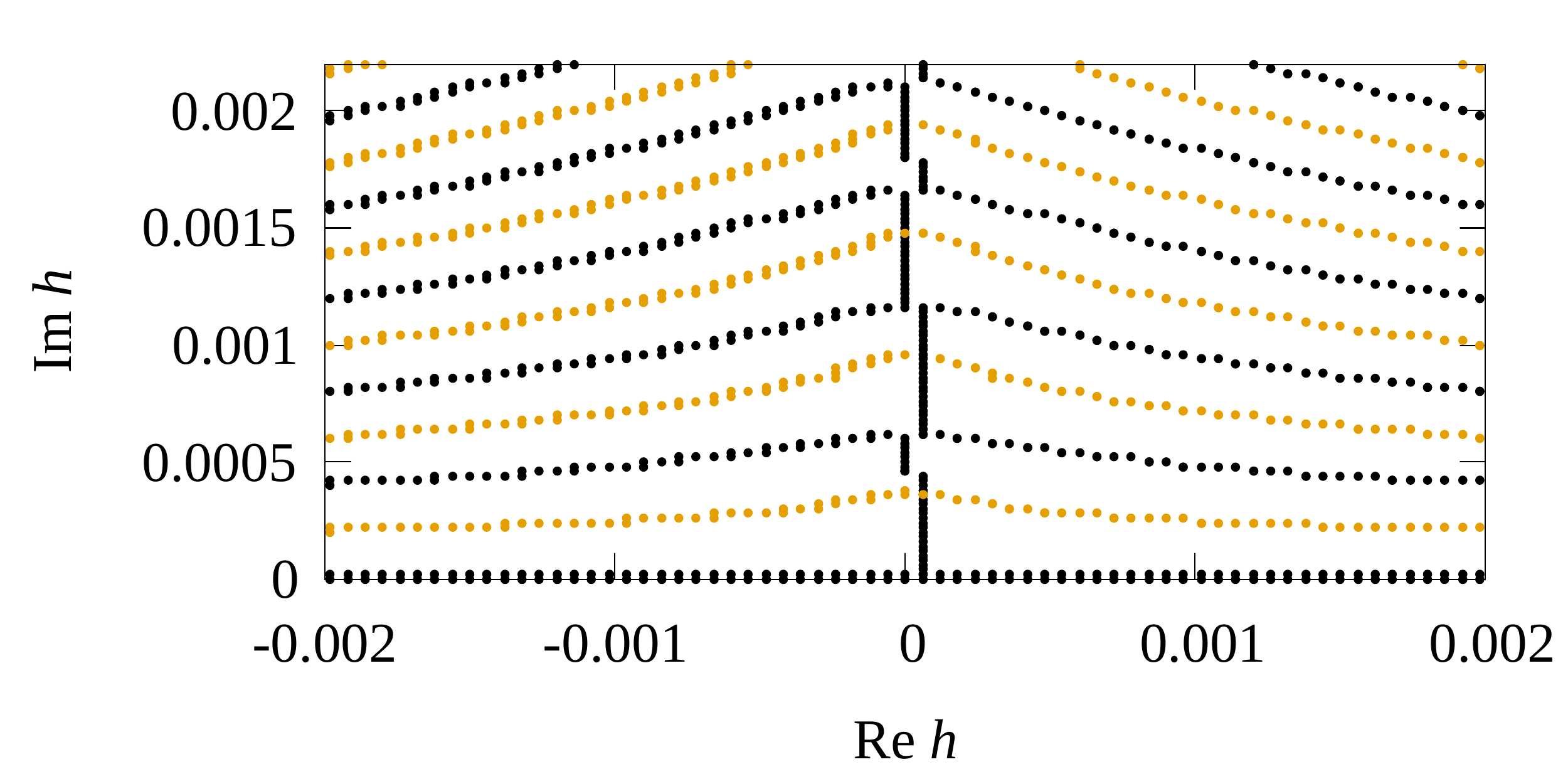}}
    	\subfloat[]{\includegraphics[width=0.5\linewidth]{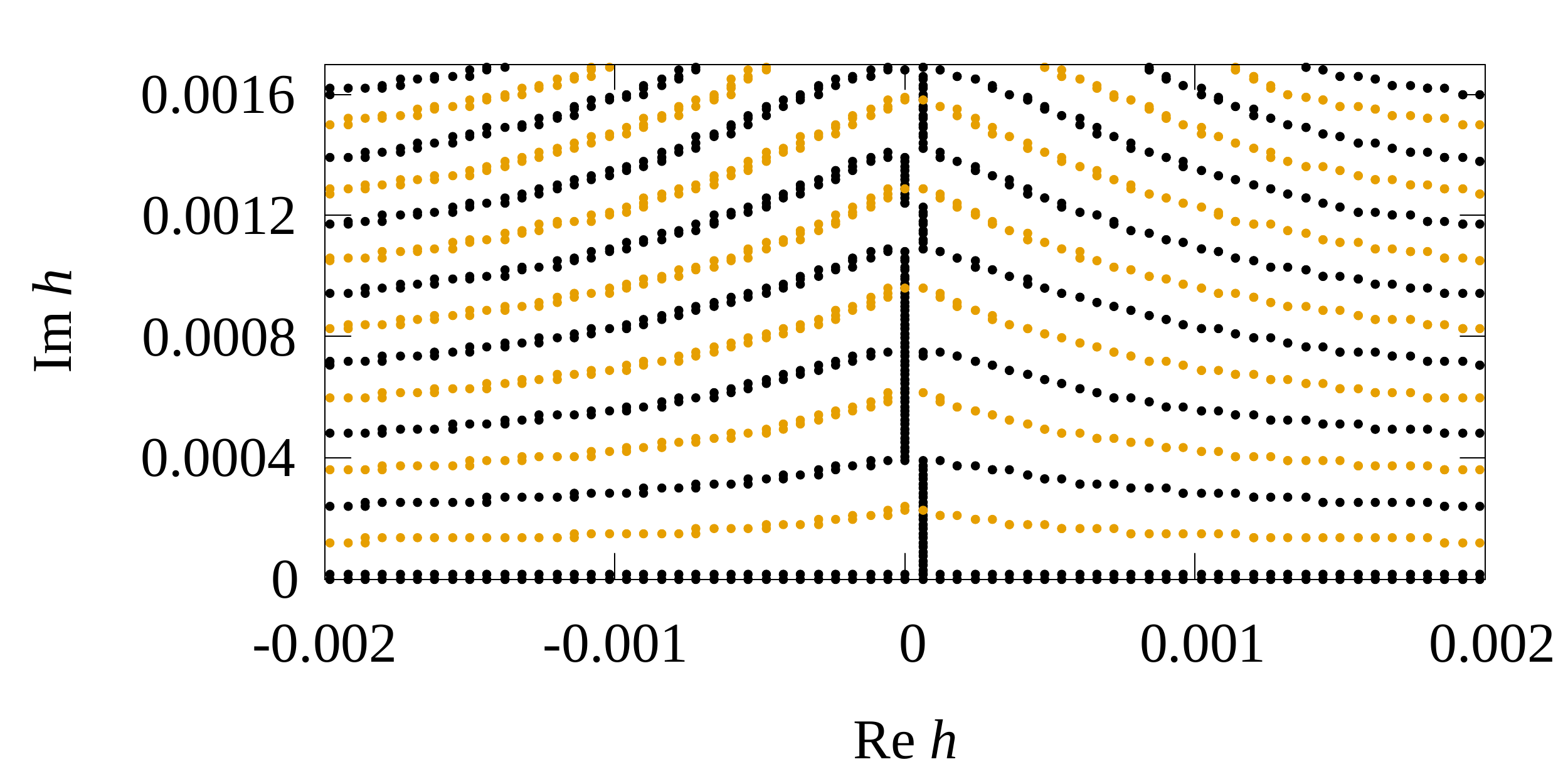}}
    	\caption{(Colour online) Lines of zeros ${\rm Im}\ \!Z(h^R, h^I) = 0$ (black curves) and ${\rm Re\ \!}Z(h^R, h^I) = 0$ (yellow curves) at the pseudo-critical temperature $T_L$ for lattice sizes $L=12,\,14,\,16,\,18$ (a, b, c, d, respectively). 
     Crossings of different color lines give the position of Lee--Yang zeros.  All zeros  are located on the imaginary axis, in agreement with the unit circle theorem.} 
    	\label{fig:lytl}
    \end{figure}
    
Let us write down the definition of the partition function in the complex $h=h^R+\ri h^I$ plane 
at $\beta=\beta_c$ more in detail:
 \begin{equation}\label{A5}
Z \big(\beta_c, h^R+\ri h^I\big)= \sum_{\{s\}} \re^{-\beta_c\sum_{i,j}S_iS_j + \beta_c h^R\sum_iS_i}
\Big[ \cos\Big(\beta_c h^I \sum_i S_i\Big) + \ri  \sin\Big(\beta_c h^I \sum_i S_i\Big) \Big]\, ,
\end{equation}
where the first sum spans all spin configurations and $S_i = \pm 1$ are Ising spins.
This can be conveniently rewritten as:
\begin{equation}\label{A6}
Z \big(\beta_c, h^R+\ri h^I\big)= Z \big(\beta_c, h^R\big) \Big[ \Big \langle \cos\Big(\beta_c h^I \sum_i S_i\Big) \Big \rangle _{\beta_c,h^R} 
+ \ri 
\Big \langle \sin\Big(\beta_c h^I \sum_i S_i\Big) \Big \rangle _{\beta_c,h^R} \Big].
\end{equation}
Here, the averaging $\left<\dots\right>_{\beta_c,h^R}$ is performed at the fixed values of temperature $\beta_c$ and magnetic field $h^R$.
 
With the data of numerical simulations at hand, to determine the points where the partition function becomes zero, it is necessary for both the real and imaginary parts of the partition function to be zero. So we look at the values of trigonometric functions averages $ \langle \cos(\beta_c h^I \sum_i S_i) \rangle_{\beta_c,h^R} $ and  $\langle \sin(\beta_c h^I \sum_i S_i) \rangle _{\beta_c,h^R}$ by plotting the contours in the complex magnetic field plane along which these functions separately vanish~\cite{contour-1,contour-2}, and then,  more precise methods are implemented around the intersection of the mentioned contours \cite{amoeba}. The simulations at the complex $h$ values are not needed as we use the histogram reweighting technique around the real simulations at $h=0$. Since the probability of states ``up'' and ``down'' are the same for the Ising model, we artificially improve our statistics by not only using the simulation data but also applying random signs to the magnetization of each state with the probability $50\%$. This way our statistics include more possible states and improve drastically.

    Results  presented in figure~\ref{fig:lytl} at $T_L$ show the position of several first Lee--Yang zeros for different lattice sizes. 
    The zeros follow a unit circle theorem, which in our coordinates means that all zeros should lie on the imaginary axis. This is what we observe for a few first zeros, though it is obscured later by the extrapolation from the simulation point. 
    
\begin{figure}
        \centering
        \includegraphics[width=0.6\linewidth]{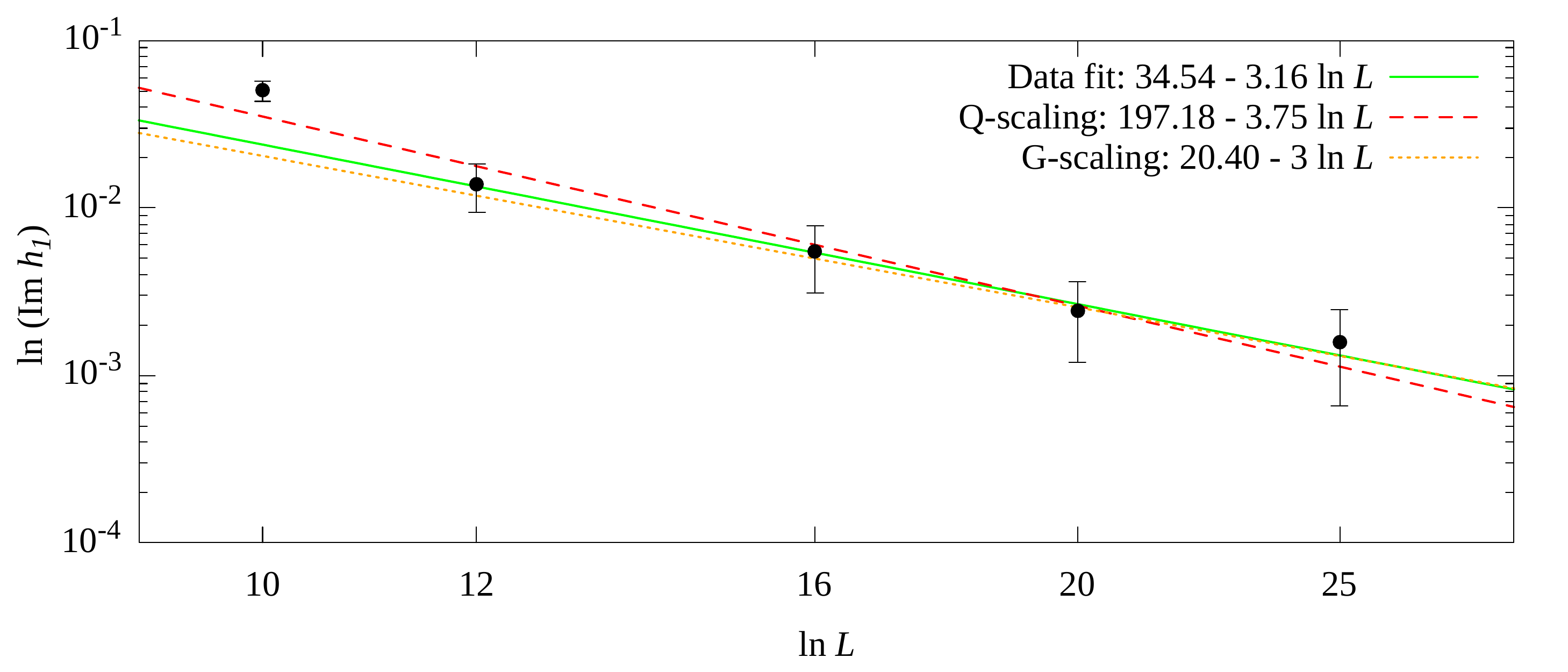}
        \caption{(Colour online) FSS of the first Lee--Yang zero for sizes $L=10,...,25$ at $T_c$. 
        Expected outcome: G-scaling. }
        \label{fig:ly_tc}
    \end{figure}

\begin{figure}
        \centering
    	\includegraphics[width=0.6\linewidth]{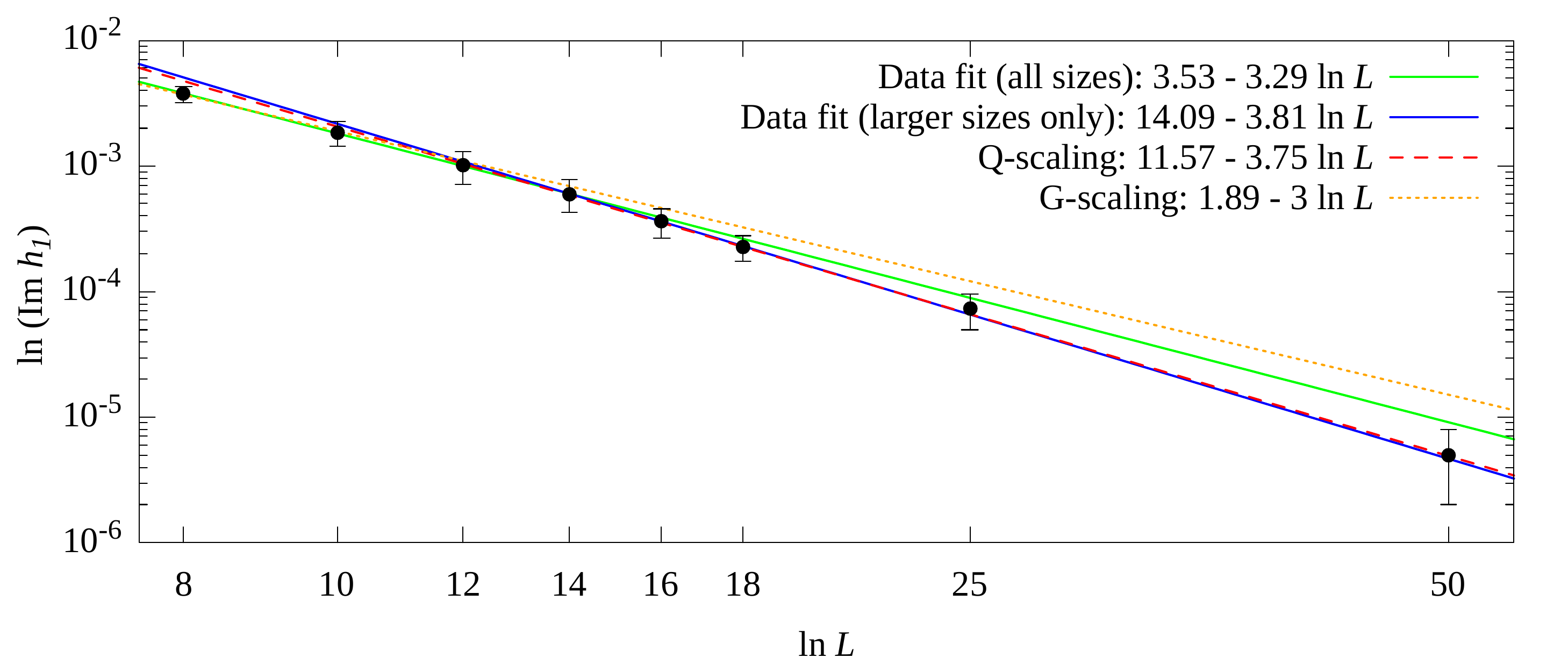}
    	\caption{(Colour online) FSS of the first Lee--Yang zero $h_1$ at  $T_L$. 
     The expected outcome: Q-scaling. We observe that for smaller sizes the results have corrections while for larger lattice sizes, the scaling comes in line with Q-scaling.}
    	\label{fig:ly_tL_fss}
    \end{figure}

    Nonetheless, we can be sure that our analysis of the first zeros is valid.  To further show this, we present a FSS of the first Lee--Yang zero in figures~\ref{fig:ly_tc}, \ref{fig:ly_tL_fss}. The FSS of {{first}} Lee--Yang zero is governed by~\cite{Itzykson}:

\be h_1 \sim L^{-\frac{\Delta}{\nu }} = L^{-\frac{\beta + \gamma}{\nu}}. 
\label{eq:LY_FSS} 
\ee
Q and G-scaling predict $h_1 \sim L^{3.75}$ and $L^3$, respectively.

\subparagraph*{Criticality:} 

 Figure~\ref{fig:ly_tc} shows the FSS of the first Lee--Yang zero at $T_c$, where we get $h_1\sim L^{-3.16 \pm 0.48}$, which might hint at G-scaling according to our other exponents at $T_c$. \par

\subparagraph*{Pseudo-criticality:}

    For the first zero $h_1$ we see that for $T_L$ our fit {{when we discard the smallest lattice sizes, $h_1\sim L^{-3.81 \pm 0.08}\,$, is well compared to Q-scaling and far from G-scaling:
$h^{(Q)}_1\sim L^{-3.75}$ versus $h^{(G)}_1\sim L^{-3}$, cf. equation~(\ref{eq:LY_FSS}).}}

{{
Furthermore, from equation~(\ref{fund}), we may write the susceptibility as
\be
 \chi \propto \frac{1}{L^d}
  \sum_j{\frac{1}{[z-z_j(L)]^2}} ,
\ee
where 
$z_j(L)$ is the $j$-th zero for a system of size $L$. 
Assuming that the first zeros dominate, this gives 
\be
 \chi \sim L^{-d} h_1^{-2} \sim L^{-d+2\Delta/\nu}.
\ee
Thus, we can provide an estimate for the susceptibility in this way.
We find the contribution of the first zero to the susceptibility, $L^{-d}h_1^{-2}\sim L^{2.62 \pm 0.10}$ which is far closer to  Q ($\chi_L \sim L^{5/2}$) than to G ($\chi_L \sim L^{2}$). 
This means that, at least from the Lee--Yang-zeros analysis, even the susceptibility is supportive of Q for pseudo-criticality and free boundaries.}}

	\section{Conclusions}

 Mean-field theory holds away for infinite volume systems above the upper critical dimension. 
Standard FSS and standard hyperscaling both fail there. 
Instead, Q theory governs finite-size systems in high dimensions.
This accounts for different types of boundary conditions and whether or not one sits at the critical or pseudo-critical point. 
Universality resides at the pseudo-critical point where standard FSS is replaced by QFSS. 
FSS at the critical point depends on boundaries. For periodic boundaries, QFSS again applies. For free boundaries, standard FSS applied to Gaussian or G modes --- not Landau MFT  --- holds (although Landau MFT is supported by simulations in the case of percolation with FBC at the percolation threshold~\cite{percolation3}). 
Here, we have used Monte Carlo simulations of the $d=5$ Ising model to verify this picture for free boundaries.

	In this paper,  we looked at the FSS of the coordinates of the pseudo-critical temperature $T_L$, mean magnetization $m$, isothermal susceptibility {{and other observables}}. 
    For the scaling of the pseudo-critical point,  we obtained $t_L \sim L^{-2}$, as predicted before {{Q}} theory in reference~\cite{Rudnick}. For mean magnetization at the critical temperature $T_c$, we observe G-scaling, while at the pseudo-critical temperature $T_L$, the small lattice sizes  {{distort}} the scaling picture, though discarding a few smallest lattice sizes gives us Q-scaling. 
    For the isothermal susceptibility, the lattice sizes in question are too small to present a proper five-dimensional scaling, and the scaling is Gaussian at both $T_c$ and $T_L$. Furthermore, the magnetization Fourier modes were studied, since for {{FBCs}} only odd magnetization Fourier modes contribute to Q-scaling. We observed that even for small lattice sizes odd modes manifest Q-scaling at the $T_L$.

 Finally, we observed that expressing susceptibility in terms of Lee--Yang zeros delivers a very strong signal of Q rather than G (or standard Landau) scaling.
 To conclude, then, all of the observables measured are supportive of the Q picture and, although small lattices cast dubious results, twelve years after our original paper \cite{pbc3}, the overall holistic picture is supportive of Q.

\section*{Acknowledgements}
We thank the Editors of this issue, Viktoria Blavatska, Taras Patsahan and Orest Pizio, for the invitation to submit this paper to the Festschrift on the occasion of the 60th anniversary of Jaroslav Ilnytskyi. Doing so we send our best wishes to Slavko, as friends call him, and express our admiration for his scientific --- and many other --- activities,
some of which we had the privilege to participate in.
We are grateful for the regular fruitful discussions among the participants of the $\mathbb{L}^4$ Collaboration and 
Doctoral College for the Statistical Physics of Complex Systems, especially to Le\"ila Moueddene, Andy Manapany, and Marjana Krasnytska. 
This work was supported in part by the National Academy of Sciences of Ukraine, project KPKBK 6541030(YuH \& YuH).
The numerical simulations were performed at Coventry University's HPC.
Our work would not be possible without the  tireless and noble  work of the Armed Forces of Ukraine.

\vspace{2ex}

 {\emph{Note added in proof.}}  One of the Authors of this paper, Ralph Kenna passed away
when this paper was in the course of editorial processing. This will be probably the
first paper we have to finalize without him, recalling again and again our common life.




\ukrainianpart

\title{Коли кореляції перевищують розмір системи: скінченновимірний скейлінґ за вільних граничних умов над верхньою критичною вимірністю}
\author{{Ю. Гончар}\refaddr{inst1,inst2,inst4}, {Б. Берш}\refaddr{inst3,inst4}, {Ю. Головач}\refaddr{inst1,inst2,inst4,inst5}, \framebox{Р. Кенна}\refaddr{inst2,inst4}}
\addresses{
	\addr{l1}{Інститут фізики конденсованих систем НАН України, 79011 Львiв, Україна}
 	\addr{l2}{Центр плинних і складних систем, Університет Ковентрі, Ковентрі, CV1 5FB, Велика Британія}
    \addr{l3}{Лабораторія теоретичної фізики та хімії, Університет Лотарингії, CNRS, Нансі, Франція}
	\addr{l4}{Співпраця $\mathbb{L}^4$ і Коледж докторантів ``Статистична фізика складних систем'', Ляйпціґ-Лотарингія-Львів-Ковентрі, Європа}
    \addr{l5}{Центр науки про складність, Відень, 1080, Австрія}
}

%
%
%

\makeukrtitle

\begin{abstract}
Ми досліджуємо скінченновимірний скейлінґ в системах з вільними граничними умовами над їх верхньою критичною вимірністю, де в термодинамічній границі критичний скейлінґ описується середньопольовою теорією.
Нещодавні роботи показують, що кореляційна довжина не обмежується фізичним розміром системи, як довгий час вважалося раніше.
Замість цього спостерігаються два режими скейлінґу --- при критичних та псевдокритичних температурах.
Ми демонструємо, що обидва режими проявляються для вільних границь.
Ми використовуємо числові симуляції моделі Ізінга при $d=5$, щоб проаналізувати намагніченість, сприйнятливість, магнітні моди Фур'є-перетворень намагніченості та нулі статистичної суми.
Хоча деякі з функцій відгуку приховують подвійний скінченовимірний скейлінґ, точність, забезпечена аналізом нулів Лі-Янга, дозволяє вивести це питання на передній план.
Зокрема, скінченновимірний скейлінґ перших нулів у псевдокритичній точці підтверджує нещодавні прогнози, що виникають з кореляцій, які перевищують розмір системи.
Ця стаття присвячена Ярославу Ільницькому з нагоди його 60-річчя.

\end{abstract}

\end{document}